\documentclass[twoside,epsf]{article}
\usepackage{graphicx}
\newcommand{\lascia}[1]{}

\newcount\Mac  \Mac=0  
\newcommand{\ifMac}[2]{\ifnum\Mac=1 #1 \else #2 \fi}

\def\Red  {}

\def\Black{}
\def\Blue {}
\newcommand{\hu}{h_{\rm u}}
\newcommand{\hd}{h_{\rm d}}

\newcommand{\mSUSY}{m_{\rm S}}
\newcommand{\STr}{{\rm STr}\,}
\newcommand{\muu}{\mu_{\rm u}}
\newcommand{\mud}{\mu_{\rm d}}
\newcommand{\muud}{\mu_{\rm ud}}

\newcommand{\BR}{\mathop{\rm BR}}

\newcommand{\GeV}{\,{\rm GeV}}

\newcommand{\One}{\hbox{1\kern-.24em I}}
\newcommand{\NP}{Nucl. Phys.}
\newcommand{\PRL}{Phys. Rev. Lett.}
\newcommand{\PL}{Phys. Lett.}
\newcommand{\PR}{Phys. Rev.}

\newcommand{\mtR}{m_{\tilde{t}_R}}

\newcommand{\mQt}{m_{\tilde{Q}_3}}
\newcommand{\eq}[1]{~(\ref{eq:#1})}

\def\circa#1{\,\raise.3ex\hbox{$#1$\kern-.75em\lower1ex\hbox{$\sim$}}\,}
\makeatletter
\def\art{\@ifnextchar[{\eart}{\oart}}
\def\eart[#1]#2#3#4#5#6{{\rm #2}, {\em #3 \bf #4} {\rm (#6) #5} ({\em #1})}
\def\hepart[#1]#2{{\rm #2, \em#1}}
\newcommand{\oart}[5]{{\rm #1}, {\em #2 \bf #3} {\rm (#5) #4}}

\newcounter{alphaequation}[equation]
\def\thealphaequation{\theequation\hbox to
0.6em{\hfil\alph{alphaequation}\hfil}}
\def\eqnsystem#1{
\def\@eqnnum{{\rm (\thealphaequation)}}
\def\@@eqncr{\let\@tempa\relax \ifcase\@eqcnt \def\@tempa{& & &} \or
  \def\@tempa{& &}\or \def\@tempa{&}\fi\@tempa
  \if@eqnsw\@eqnnum\refstepcounter{alphaequation}\fi
\global\@eqnswtrue\global\@eqcnt=0\cr}
\refstepcounter{equation} \let\@currentlabel\theequation \def\@tempb{#1}
\ifx\@tempb\empty\else\label{#1}\fi
\refstepcounter{alphaequation}
\let\@currentlabel\thealphaequation
\global\@eqnswtrue\global\@eqcnt=0 \tabskip\@centering\let\\=\@eqncr
$$\halign to \displaywidth\bgroup \@eqnsel\hskip\@centering
$\displaystyle\tabskip\z@{##}$&\global\@eqcnt\@ne
\hskip2\arraycolsep\hfil${##}$\hfil& \global\@eqcnt\tw@\hskip2\arraycolsep
$\displaystyle\tabskip\z@{##}$\hfil
\tabskip\@centering&\llap{##}\tabskip\z@\cr}
\def\endeqnsystem{\@@eqncr\egroup$$\global\@ignoretrue} \makeatother

\newcount\n

\begin{document}\twocolumn[
\centerline{hep-ph/0005203 \hfill IFUP--TH/2000--14 \hfill SNS-PH/00--10}
\vspace{5mm}

\Black
\vspace{0.5cm}
\centerline{\LARGE\bf\Red Relating $m_{\rm Higgs}$ to $m_{\rm SUSY}$ by a loop factor}
\medskip\bigskip\Black
\centerline{\large\bf Riccardo Barbieri}\vspace{0.2cm}
\centerline{\em Scuola Normale Superiore, Piazza dei Cavalieri 7, I-56126 Pisa, Italy and INFN}
\vspace{3mm}
\centerline{\large\bf Alessandro Strumia}\vspace{0.2cm}
\centerline{\em Dipartimento di Fisica, Universit\`a di Pisa and INFN, Pisa, Italia}
\vspace{1cm}
\Blue
\centerline{\large\bf Abstract}
\begin{quote}\large\indent
We assume that the scale of the soft supersymmetry breaking masses, $\mSUSY$, sliding at tree level,
is fixed by the minimization of the potential, but without a significant contribution of the cosmological term.
Under these assumptions the electroweak breaking scale gets related to $\mSUSY$ by a loop factor.
Applied to specific models of supersymmetry breaking,
this mechanism naturally gives sparticles slightly above all present
accelerator bounds.

\end{quote}\Black
\vspace{0.5cm}]

\noindent\paragraph{1}
The search for supersymmetry has not been successful so far.
No superpartner has been found yet. At the same time, the lower limit on the mass of the Higgs field, $m_h$,
is steadily growing and already covers a large portion of the range expected for the
mass of the lightest Higgs scalar in the Minimal Supersymmetric Standard Model (MSSM).

In some way, this is in contrast with the spectacular success of the correlation foreseen in a unified
supersymmetric theory between the strong coupling constant $\alpha_3$ and 
the weak mixing angle~\cite{GUT}.
Similarly, the significant direct lower limit on the Higgs mass must be contrasted with the range for the same
mass indirectly indicated by the electroweak precision tests~\cite{LEP}, centered on the expectation in a
supersymmetric extension of the SM.

We have to be more precise here. As well known, the successful prediction for $\alpha_3(M_Z)$
in a unified supersymmetric theory in no way requires superpartners so light that they had to be already discovered.
On the contrary, if one could neglect uncertainties in this prediction associated for example with the
threshold effects at unification or with the nearby scale of gravity, the central value of the prediction,
$\alpha_3(M_Z)=0.128(10)$~\cite{a3GUT}, slightly higher than the observed value,
$\alpha_3(M_Z)_{\overline{\rm MS}}=0.119(3)$, might even indicate superpartners heavier than those
accessible to present day searches.

The problem, rather, is of theoretical nature and is related to the fine-tuning argument.
In existing supersymmetric models, all superpartner masses can be pushed above the present
experimental limits only at the price of some level of apparently accidental 
tuning among different parameters~\cite{FT}.
While such tuning for sure cannot be excluded, it would be better to have an explanation for it.
Such an explanation would be particularly welcome if, at the same time, it shed some light on the same
actual values of superparticle masses.

The problem in existing models is the occurrence of at least one of the
following two instances.
Without fine-tuning, either the underlying physics sets, at lowest order, the Higgs mass parameters at values
similar to $\mSUSY$, the typical supersymmetric mass,
or the radiative corrections to the Higgs masses involve large logarithms and make them again comparable
to $\mSUSY$ (or more precisely to the gluino mass).
Hence an expectation for $\mSUSY$ in the same range of $M_Z$, taking into account the gauge nature of the
quartic couplings of the MSSM Higgs fields.
In our view, to find a model which avoids both these correlations between the Higgs mass parameters
and $\mSUSY$ in a natural way is a phenomenologically relevant 
and theoretically well defined problem.
In such a model $M_Z^2$ would be related to $\mSUSY^2$ by a loop factor without large logarithms, thus
relaxing the naturalness upper bounds on sparticle masses.

We do not know of a straight solution for this problem.
We can, however, present a hypothesis which, if it were satisfied in a given model, would precisely solve it.
As we shall see, the implications for some of the superpartner masses would also be
rather precise.

\paragraph{2} We assume that a theory exists where the scale of the soft supersymmetry breaking masses is
not fixed at the classical level but is rather determined by the quantum corrections associated with the
light MSSM fields~\cite{slide}.
Furthermore we assume that the cosmological term in the effective potential, to be precisely
defined in a while, does not influence in a significant way such determination.

A specific implementation of this hypothesis could be as follows.
All soft supersymmetry breaking masses are proportional to a unique dynamical field $\mSUSY$, which
parameterizes a flat direction of the potential at the classical level.
This $\mSUSY$ field must be viewed as an extra dynamical variable in addition to the usual MSSM
Higgs fields $\hu$ and $\hd$~\cite{slide}.
Furthermore in the effective potential $V_{\rm eff}(\hu,\hd,\mSUSY)$, the
``cosmological term'' $V_{\rm eff}(0,0,\mSUSY)$ is either identically vanishing or is small enough not
to influence the determination of $\mSUSY$.

In general $V_{\rm eff}(0,0,\mSUSY)$, which receives contributions from all particles coupled to $\mSUSY$,
is expected to contain both a quadratic and a logarithmic term in the cut-off.
At one loop these terms are respectively proportional to
$\STr M^2$ and $\STr M^4$, where $M$ is the general, $\mSUSY$-dependent mass matrix in the theory.
We can offer no explanation of why these terms, or those occurring in higher order, should cancel
out to a sufficient degree not to play any significant role in the determination of $\mSUSY$, but this is
what we assume.
This might perhaps have something to do with the problem of the cosmological constant.

The point of this paper is that, if this hypothesis can be met, not only  $\mSUSY$ is fixed
at an exponentially small value with respect to the cut-off but also the problem formulated
in the previous section is neatly solved.

In words, the idea can be simply explained as follows.
By the usual Coleman-Weinberg mechanism the relevant Higgs squared mass, which is born positive
at the cut-off $\Lambda$, is led to zero in the infrared at a scale $Q$ exponentially far from
$\Lambda$~\cite{RGEMSSM}.
If $\mSUSY$ were fixed, $Q$ would have to be bigger than $\mSUSY$ but would otherwise be unrelated to it.
On the contrary, if $\mSUSY$ is a dynamical field, the minimization of the potential wants $\mSUSY$ to get
close to $Q$.
This is precisely what one is looking for.
In this way $M_Z$ gets related to $\mSUSY$ by a loop factor without large logarithms.

\paragraph{3} The analytic implementation of this idea is as follows.
The MSSM  Higgs potential for the neutral Higgs fields is
\begin{equation}\label{eq:VMSSM}
V=\muu^2\hu^2+\mud^2\hd^2-2\muud^2\hu\hd + \frac{\bar{g}^2}{8}(\hu^2-\hd^2)^2
\end{equation}
where $\bar{g}^2=g_Y^2+g_2^2$ and the $\mu_i^2$ must be viewed as proportional to $\mSUSY^2$,
like all other mass parameters in the Lagrangian.
A further logarithmic dependence on $\mSUSY$ of the $\mu_i^2$ is induced by the usual running.

Minimizing $V$ with respect to the Higgs fields gives
$V_{\rm min}= -{\mu_{\rm eff}^4}/{2\bar{g}^2}$
where
\begin{equation}\label{eq:defs}
\sin2\beta=\frac{2\muud^2}{\muu^2+\mud^2},\qquad
\bar{g}^2 v^2\cos2\beta=2\mu_{\rm eff}^2
\end{equation}
and
\begin{equation}
\mu_{\rm eff}^2 = \muu^2-\mud^2+\sqrt{(\muu^2+\mud^2)^2-4\muud^4}.
\end{equation}
Having assumed that no other $\mSUSY$ dependent term contributes significantly to $V$, the
``sliding scale'' $\mSUSY$ is determined by the minimization of $V_{\rm min}$.
Neglecting the running of $\bar{g}$, which becomes formally a two loop 
effect, we obtain
\begin{equation}
\mSUSY\frac{\partial V_{\rm min}}{\partial\mSUSY} = 
-\frac{\mu_{\rm eff}^2}{2\bar{g}^2}(4\mu_{\rm eff}^2+2\dot{\mu}_{\rm 
eff}^2),
\end{equation}
where the dot stands for $d/d\ln\bar{\mu}$ and $\bar{\mu}$ is the running scale.
An explicit calculation gives
\begin{equation}
\dot{\mu}_{\rm eff}^2 = -\frac{2}{\cos2\beta}\big
[\dot{\mu}_{\rm u}^2 \sin^2\beta+\dot{\mu}_{\rm d}^2\cos^2\beta-2\dot{\mu}_{\rm ud}^2\sin\beta\cos\beta\big]
\end{equation}
so that, using\eq{defs} and $M_Z^2=\bar{g}^2 v^2/2$, we get
\begin{equation}\label{eq:minmin}
\Blue M_Z^2 \cos^2 2 \beta = 
\dot{\mu}_{\rm u}^2\sin^2 \beta  + \dot{\mu}_{\rm d}^2 \cos^2\beta -2\dot{\mu}_{\rm ud}^2 \sin\beta \cos\beta  \Black.    
\end{equation}
Upon replacement of the $\dot{\mu}_i^2$ with their explicit expressions, this is a relation between $M_Z^2$ and the
soft supersymmetry breaking masses
involving a loop factor as promised.
This relation supplements the standard formula
\begin{equation}\label{eq:min}
{M_Z^2\over 2} =\frac{\mud^2-\muu^2\tan^2\beta}{\tan^2\beta-1}
\end{equation}
and implies an apparent fine tuning of order $(4\pi)^2$ in it.

Eq.\eq{minmin} is a lowest order formula.
A dominant correction to it occurs via the effects of the top/stop exchanges on the quartic
Higgs couplings, as in the case of the Higgs masses~\cite{mhloop}.
As well known, for $\mSUSY\gg M_Z$, only one Higgs field $h$ remains light from\eq{VMSSM},
approximately given by $h=\hd\cos\beta+\hu\sin\beta$. The potential
in this only field $h$ is
$V(h) = m^2 h^2 + \lambda h^4$,
where
$$m^2 = \muu^2\sin^2\beta+\mud^2\cos^2\beta-2\muud^2\sin\beta\cos\beta$$
and
$$\lambda = \frac{\bar{g}^2}{8}\cos^22\beta + \frac{6}{(4\pi)^2}\lambda_t^4\ln\frac{m_{\rm SUSY}}{h}$$
which also includes the top/stop effects.
Minimizing $V(h)$, as before, with respect to $h$ and to $\mSUSY$ leads to the same equation\eq{minmin}
with the replacement $M_Z^2\cos^2 2\beta\to m_h^2$,
where $m_h$ coincides with the lightest Higgs mass in the approximation that we are considering.

\begin{figure}[t]
\begin{center}
\includegraphics{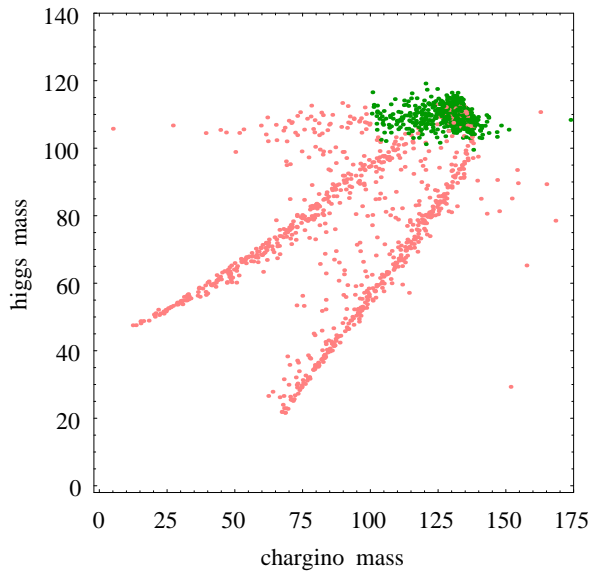}
\caption[SP]{\em Correlation between the chargino and the Higgs mass in minimal supergravity.
Sampling spectra excluded (allowed) by accelerator bounds are drawn in light red (dark green).
The two branches arise from opposite signs of $\mu$.
\label{fig:FT}}
\end{center}\end{figure}

\paragraph{4}
Let us now discuss the consequences of our assumptions.
For simplicity we will present them in the limit of moderately large 
$\tan\beta\sim(5\div 10)$, so that\eq{minmin} simplifies to
$$M_Z^2 = \frac{1}{(4\pi)^2}\big[6\lambda_t^2(\mu_{\rm
u}^2+2m_{\tilde{t}}^2)-4c_i^h g_i^2 (M_i^2+\mu^2) \big]    
$$
where $2m_{\tilde{t}}^2\equiv \mQt^2+\mtR^2+A_t^2$ is a combination of stop mass parameters,
$\{c_1^h,c_2^h,c_3^h\}=\{3/10,3/2,0\}$,
and we have used standard notations for the various parameters.
The standard minimization, eq.\eq{min}, tells that $\muu^2=-M_Z^2/2$ is negligible.
If also the terms proportional to the electroweak gauge couplings can be neglected, then
$m_{\tilde{t}} \approx 4\pi M_Z/\sqrt{12}$.
Knowing the value of the stop scale, we can estimate that
top/stop loop effects increase by $\sim (20\div 30)\%$
the Higgs mass above the tree level MSSM prediction
($m_h=M_Z$ in the large $\tan\beta$ and heavy sparticle limits).
For $\tan\beta>5$, $m_h$ ranges from $110$ to $120\GeV$.
As said, these loop corrections also increase by the same amount
the prediction for $m_{\tilde{t}}$, so that $m_{\tilde{t}}\sim  (380\div 430)\GeV$.

Contrary to the case of the standard minimization\eq{min}, 
a significant cancellation in\eq{minmin}
can make sparticles unnaturally heavier than their typical values
only for quite unplausible values of the parameters.
Such cancellations do not happen in
some popular models.

In minimal supergravity~\cite{mSUGRA} (i.e.\ with universal soft terms at the unification scale)
the RGE effects that induce EW symmetry breaking below the
scale $Q$ are dominated by the strong sector.
In fact $Q$ is most naturally of order of the scale at which the strong coupling
begins to be `strong' and the gluino begins to be `heavy'.
For wide ranges of boundary conditions at the unification scale, terms proportional to the squared gluino mass
end up dominating the r.h.s.\ of\eq{minmin}, that therefore
becomes mainly a prediction for the gluino mass:
$M_3(M_3) \approx 450\GeV|\cos 2\beta|$.
For moderately large $\tan\beta$ this implies a chargino mass $m_\chi\approx 125\GeV$.
A sampling of the parameter space (see fig.~1) shows that in about
half of this space (and almost always when $\tan\beta\circa{>}4$)
the self-tuning mechanism naturally gives sparticles above all accelerator 
exclusion bounds, but not far from them.
Sparticle loop effects often produce too large corrections to $\BR(B\to X_s\gamma)$;
a significant part of the remaining portion of the parameter space should be probed by the planned
measurement of anomalous magnetic moment of the muon.
The thermal relic abundance of neutralinos~\cite{OmegaNeutr} can be in the cosmologically interesting range.

In more predictive models eq.\eq{minmin} allows to fix
the sparticle spectrum.
This happens for example in gauge mediation models~\cite{GM} if we assume that
the unknown physics that generates the $\mu$ term does not
give at the same time additional contributions to the Higgs and stop masses.
In this case the sparticle spectrum is determined
as function of $\tan\beta$, of the effective number of messengers $n$ and of the 
mediation scale $M$.
For moderately large $\tan\beta$
the right-handed sleptons are heavier than $100\GeV$ when $M\circa{>}10^{10}\GeV$.
In this region the gluino mass depends mainly on $n$ and ranges between 350 and $430\GeV$.

Finally, a general consequence of our assumptions is the presence in $\mSUSY$ of at least one scalar field.
Depending on its unknown kinetic term, it could either be 
a light scalar with small couplings, constrained by `fifth force' experiments,
or, in the opposite limit, it could manifest itself as a state with a weak scale mass,
that contributes to the invisible width of the Higgs
and sligthly mixes with it.


\paragraph{Acknowledgments}
Work supported in part by the E.C.\ under TMR contract No. ERBFMRX--CT96--0090.
We thank Riccardo Rattazzi for useful discussions.


\frenchspacing
\small\footnotesize
\begin{thebibliography}{nn}
\bibitem{GUT}
\art{S. Dimopoulos, S. Raby, F. Wilckzek}{\PR}{D24}{1681}{1981};
\art{L.E. Ib\'a\~nez, G.G. Ross}{\PL}{105B}{439}{1981};
\art{S. Dimopoulos, H. Georgi}{\NP}{B193}{375}{1981};



\bibitem{LEP} LEP working group.
Avaible  at the www address
{\tt www.web.cern.ch/LEPEWWG}.
The full set of data involves measurements at LEP, SLC and Tevatron.


\bibitem{a3GUT}
\art[hep-ph/9503214]{P. Langacker, N. Polonsky}{\PR}{D52}{3081}{1995}.


\bibitem{FT}
\art{R. Barbieri, G.F. Giudice}{\NP}{B306}{63}{1988};
\hepart[hep-ph/9904247]{A. Strumia} and ref.s therein.

\bibitem{slide}
For attempts to build models with similar properties, see
\art{E. Cremmer et al.}{\PL}{B133}{61}{1983};
\art{J. Ellis, A.B. Lahanas, D.V. Nanopoulos, K. Tamvakis}{\PL}{B134}{429}{1984}.
See also
\art[hep-ph/9406256]{C. Kounnas, F. Zwirner, I. Pavel}{\PL}{B335}{403}{1994}.
See
\art[hep-ph/9504296]{S. Dimopoulos, G.F. Giudice, N. Tetradis}{\NP}{B454}{59}{1995}
for a possible generalization to flavour-breaking parameters.


\bibitem{RGEMSSM}
\art{L. Ib\'a\~nez, G.G. Ross}{\PL}{B110}{215}{1982};
\art{L. Ib\'a\~nez, C. Lopez}{\NP}{B233}{511}{1984};
\art{A. Bouquet, J. Kaplan, C.A. Savoy}{\NP}{B262}{299}{1985}.


\bibitem{mhloop}
\art{H.E. Haber, R. Hempfling}{\PRL}{66}{1815}{1991};
\art{J. Ellis, G. Ridolfi, F. Zwirner}{\PL}{275B}{83}{1991};
\art{Y. Okada, M. Yamaguchi, T. Yanagida}{Prog. Theor. Phys. Lett.}{85}{1}{1991};
\art{R. Barbieri, F. Caravaglios, M. Frigeni}{\PL}{258B}{167}{1991}.

\bibitem{mSUGRA}
\art{R. Barbieri, S. Ferrara, C. Savoy}{\PL}{119B}{343}{1982};
\art{P. Nath, R. Arnowitt, A. Chamseddine}{\PRL}{49}{970}{1982}.

\bibitem{OmegaNeutr}
See e.g.
\art{J. Ellis, J.S. Hagelin, D.V. Nanopoulos, K.A. Olive, M. Srednicki}{\NP}{B238}{453}{1984};
\art{M. Drees, M. Nojiri}{\PR}{47}{376}{1993}.

\bibitem{GM}
\art{M. Dine, W. Fishler, M. Srednicki}{\NP}{B189}{575}{1981};
\art{S. Dimopoulos, S. Raby}{\NP}{B192}{353}{1981};
\art{L. Alvarez-Gaume, M. Claudson, M.B. Wise}{\NP}{B207}{96}{1982}.

\end{thebibliography}
\end{document}